\documentclass[letterpaper,twocolumn,10pt]{article}
\usepackage{usenix-2020-09}

\usepackage{tikz}
\usepackage{amsmath}
\usepackage{adjustbox}

\usepackage{filecontents}

\usepackage{microtype}
\usepackage{subcaption}
\usepackage{graphicx}
\usepackage{booktabs} 

\usepackage{hyperref}
\usepackage{xcolor}
\usepackage[linesnumbered,ruled,noend]{algorithm2e}
\usepackage{multirow}
\usepackage{listings}
\usepackage{epstopdf}
\usepackage{amssymb}
\usepackage{enumitem}
\usepackage[frozencache,cachedir=minted-cache]{minted} 
\usepackage{balance}
\usepackage{cleveref}
\usepackage{array}
\usepackage[normalem]{ulem}
\usepackage{tcolorbox}
\usetikzlibrary{positioning}
\usepackage{listings}
\usepackage{threeparttable}
\setlist{leftmargin=*}
\newcommand{\sys}{{\textsc{Jenga}}\xspace}

\newcommand{\tofill}[1]{\cred{$\times\times\times$}\xspace}
\definecolor{darkgreen}{rgb}{0,0.8,0}
\definecolor{codebg}{rgb}{0.2,1,0.7}
\newcommand{\cred}{\textcolor{red}}

\newcommand{\cgreen}{\textcolor{darkgreen}}

\definecolor{teal}{RGB}{55, 149, 189}

\newcommand{\latestvllm}{vLLM v0.6.4\xspace}

\newcommand{\para}[1]{\noindent \textbf{#1 }}

\setminted[python]{linenos,frame=single,breaklines,fontsize=\footnotesize, escapeinside=@@}

\newcommand{\llama}{Llama\xspace}
\newcommand{\gemma}{Gemma-2\xspace}

\crefformat{section}{\S#2#1#3}
\crefformat{subsection}{\S#2#1#3}
\crefrangeformat{section}{\S#2#1#3}
\crefrangeformat{subsection}{\S#2#1#3}
\setlist[itemize]{noitemsep, topsep=0pt}
\setlist[enumerate]{noitemsep, topsep=0pt}
\crefformat{figure}{Figure #2#1#3}
\crefformat{subfigure}{Figure #2#1#3}
\crefformat{table}{Table #2#1#3}
\SetArgSty{textit}{}{}

\newcommand{\tightsection}[1]{
\vspace{-0.1cm}
\section{#1}
\vspace{-0.1cm}
}

\newcommand{\tightsubsection}[1]{
\vspace{-0.1cm}
\subsection{#1}
\vspace{-0.1cm}
}

\begin{document}

\date{}

\title{\Large \bf \sys: Effective Memory Management for Serving LLM with Heterogeneity}

\renewcommand\footnotemark{}
\author{
{\rm
Chen Zhang$^{\dag*}$\thanks{$^{*}$Part of the work was done during visiting UC Berkeley.}, 
Kuntai Du$^{\ddag*}$,
Shu Liu$^{\diamond}$, 
Woosuk Kwon$^{\diamond}$,
Xiangxi Mo$^{\diamond}$,
Yufeng Wang$^{\S}$,
Xiaoxuan Liu$^{\diamond}$
}
\and
{\rm
Kaichao You$^{\dag*}$,
Zhuohan Li$^{\diamond}$,
Mingsheng Long$^{\dag}$,
Jidong Zhai$^{\dag}$,
Joseph Gonzalez$^{\diamond}$,
Ion Stoica$^{\diamond}$ 
}
\and
$^{\dag}$Tsinghua University \hspace{0.4cm}
$^{\ddag}$University of Chicago \hspace{0.4cm}
$^{\diamond}$UC Berkeley \hspace{0.4cm}
$^{\S}$Independent Researcher
} 

\maketitle

\begin{abstract}
Large language models (LLMs) are widely used but expensive to run, especially as inference workloads grow. To lower costs, maximizing the request batch size by managing GPU memory efficiently is crucial.
While PagedAttention has recently been proposed to improve the efficiency of memory management, we find that the growing heterogeneity 
in the embeddings dimensions, attention, and access patterns
of modern LLM architectures introduces new challenges for memory allocation.

In this paper, we present \sys, a novel memory allocation framework for heterogeneous embeddings in LLMs.
\sys tackles two key challenges: (1) minimizing memory fragmentation when managing embeddings of different sizes, and (2) enabling flexible caching and eviction policies tailored to the specific token-dependency patterns of various layers.
\sys employs a two-level memory allocator, leveraging the least common multiple (LCM) of embedding sizes to optimize memory usage and providing APIs to express layer-specific caching logic to enhance memory reuse.

We implemente \sys on vLLM, a state-of-the-art LLM inference engine, and evaluate it with diverse LLMs, datasets, and GPU configurations. Evaluations show that \sys improves GPU memory utilization by up to 79.6\%, and increases serving throughput by up to 4.92$\times$ (1.80$\times$ on average).



\end{abstract}

\section{Introduction}
\label{sec:intro}


The broad adoption of LLMs for services like 
ChatGPT~\cite{cnbc2024openai} and GitHub Copilot~\cite{ciodive2024github} 
has driven a surge in demand for GPUs to power LLM serving.
Serving LLMs requires many expensive GPUs to achieve the desired service latency objectives.
Even hyperscalers like Microsoft struggle to secure sufficient GPU capacity for their LLM-powered services~\cite{cnbc_msft}.
Because tokens are generated sequentially during inference, LLM serving workloads often fail to adequately utilize the compute capacity of modern GPUs. 

One way to reduce the cost of LLM serving is to increase GPU utilization by processing multiple requests in parallel (i.e., batching requests). 
While batching often has minimal impact on latency, batching introduces a new challenge. 
Each generated token may depend on the computed embeddings of all previous tokens in the same request.
Therefore, we need to store the embeddings for the prefixes of all requests belonging to the same batch in the GPU memory. 
This makes GPU memory \emph{capacity}
the new bottleneck to fully utilizing the GPU compute. 
Thus, efficient memory management is the key to improving the GPU throughput, and consequently alleviating the high costs of LLM serving.

By borrowing ideas from memory management in operating systems, PagedAttention~\cite{kwon2023efficient} introduced mechanisms to reduce fragmentation of the embeddings for all previous tokens (KV-Cache).
PagedAttention manages a mapping between virtual and physical pages.
Each page is a fixed-size multiple of an embedding, allowing for more efficient allocation and minimizing fragmentation. 
As a result, PagedAttention was able to improve throughput by 2 - 4$\times$ compared to previous state-of-the-art solutions~\cite{yu2022orca}, and is now used by virtually all LLM inference engines, including vLLM~\cite{kwon2023efficient}, SGLang~\cite{zheng2023efficiently} and TensorRT-LLM~\cite{trtllm}.

\begin{figure}
    \centering
    \includegraphics[width=0.48\textwidth]{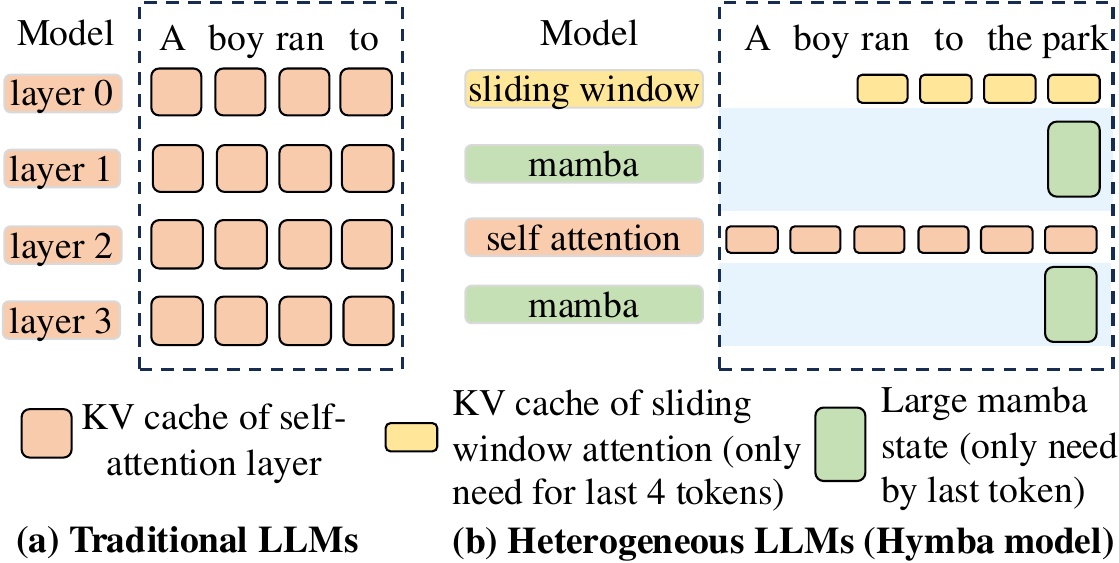}
    \caption{Traditional LLMs (left) v.s. Latest LLMs (right). LLMs are becoming more and more heterogeneous and produce KV caches with different sizes and dependencies, which demands a new GPU memory manager design.}
    \label{fig:contrast}
\end{figure}

The design of PagedAttention was built on two fundamental assumptions about LLM architectures, which held true when monolithic Transformers~\cite{vaswani2017attention} were dominant:

\begin{enumerate}
    \item \textbf{Fixed-size embeddings}: Embeddings are the same size in different tokens and layers. As such, the granularity of memory allocation is naturally a fixed page size, which is a multiple of the embedding size. 
    \item \textbf{Full-prefix dependency}: Generating the next token on a request depends on \emph{all} previous tokens of the request.
    As such, all tokens in the prefix share the same life cycle.
\end{enumerate}

Two years after the introduction of PagedAttention, LLM architectures have evolved to incorporate more \emph{heterogeneous} components (Figure~\ref{fig:contrast}), which invalidates both assumptions.




First, recent models often have heterogeneous embeddings with different sizes.
For example, vision-language models (VLMs) such as LLaVA~\cite{liu2024visual} and InternVL~\cite{chen2024internvl} retain vision embeddings for image inputs as well as KV cache for each token.
The sizes of these embeddings are naturally different as they encode different types of information.
Moreover, recent approaches~\cite{lieber2024jamba,dong2024hymba} combine Mamba~\cite{gu2023mamba} with the regular LLM architecture, where the Mamba state representing a sequence is typically larger than the KV cache associated with a single token.

Second, to handle long contexts more efficiently, some new LLM architectures use only a subset of the prefix tokens to generate the next token. We call this token-dependency pattern \textit{prefix-subset dependency}. For instance, Google's \gemma~\cite{team2024gemma} interleaves full attention with sliding window attention~\cite{beltagy2020longformer}: some layers attend to the entire prefix, while other layers only attend to a sliding window of the most recent tokens.
Building on this, NVIDIA's Hymba~\cite{dong2024hymba} further integrates Mamba layers, which solely rely on the representation of the last token. 
These architectures invalidate the second assumption that the generation of the next token depends on \emph{all} previous tokens of the query. 

As a result, the throughput of PagedAttention can drop by up to $4.91\times$  compared to an ideal solution which only stores the required tokens needed by each layer to compute the next token, and tightly packs different-size embeddings in the KV cache (see Section~\ref{sec:eval}). 

To close this gap, we propose \sys, a new memory management framework for the KV cache. There are two challenges which directly follow from the violation of the two assumptions that \sys needs to address: (1) minimize the fragmentation when handling different-size embeddings, and (2) customize memory eviction and caching policies for each type of layer to minimize cache misses.

To address the first challenge, \sys uses a two-level memory allocator. At the bottom layer, we still have fixed-size pages, while at the top layer we have different-size embeddings. To minimize internal fragmentation, we use the know model architecture to
pick a compatible page size so that it is a multiple of each embedding size. 
In particular, we choose the page size as \emph{least common multiple} (LCM) of token embedding sizes. For example, if we have embeddings of two different sizes, $2$KB and $3$KB, respectively, we pick a page with a compatible size of $6$KB. We call such an allocator 
\textit{LCM allocator}. Note that the LCM allocator can be seen as a particular case of a slab memory allocator~\cite{bonwick1994slab}, commonly used in modern operating systems. 
However, here we take advantage of the fact that we know a priori all embedding sizes to simplify the design and minimize the fragmentation.
Furthermore, \sys uses a request-aware allocation strategy to further reduce fragmentation.

To address the second challenge, \sys allows the application (in this case LLM serving) to customize eviction and caching policies for a wide range of attention layers such as sliding window, Mamba, and cross-attention.
\sys achieves this by providing a general mechanism to handle prefix-subset dependencies, and enable attention variants to easily customize this mechanism by precisely specifying the exact prefix subset required to generate the next token.

We have implemented and evaluated \sys in a wide variety of models and use cases: from heterogeneous attention layers within the same model, to KV caches of draft and target models in speculative decoding~\cite{leviathan2023fast}, and to vision embeddings of VLMs.
Compared to vLLM, a state-of-the-art LLM serving engine, \sys improves memory utilization by up to 79.6$\%$, and achieves up to $4.92\times$ higher throughput (1.80$\times$ on average) without impacting end-to-end latency.

In summary, this paper makes the following contributions.
\begin{itemize}
    \item We identify the growing heterogeneity of new LLM architectures, both in terms of embedding sizes and token dependencies to generate a new token.
    \item We propose a two-level memory allocation system for KV cache that efficiently supports different-size embeddings, and provides flexibility to customize caching and eviction policies to maximize the hit rate for different types of layers.
    \item A prototype implementation, \sys, on top of vLLM, a production-level inference engine that provides support for a wide range of optimizations and features.
    \item Achieve a 1.80$\times$ increase in throughput over state-of-the-art LLM inference engines, without any impact on latency.
\end{itemize}

\tightsection{Background}








This section introduces the basic concepts in LLM inference.


\para{KV Cache.}
LLMs are autoregressive models that generate tokens iteratively, one at a time. The inference engine must store KV caches—intermediate tensors produced by the attention layers—for each token inside GPU because The computation of a new token depends on interactions between its embedding and the previously stored intermediate KV cache tensors. 

The size of KV caches can be substantial. For instance, the KV cache size for Llama 3.1 8B~\cite{dubey2024llama} is approximately 1.2 GB for a single request with ten thousand tokens. Moreover, LLM serving systems typically batch tens of requests for efficient inference~\cite{kwon2023efficient}, requiring them to manage tens of GBs of KV caches. Therefore, the KV cache needs to be managed carefully to achieve high inference efficiency.

\para{Prefix Caching.} The KV caches can be retained in GPU memory even after the corresponding request's generation is complete. This allows subsequent requests with shared prefixes to reuse these caches, thereby reducing redundant computation. Prefix caching is particularly effective when multiple queries share a common prefix, such as when asking different questions about the same long document.



\begin{figure*}[t]
    \centering
    \includegraphics[width=0.9\textwidth]{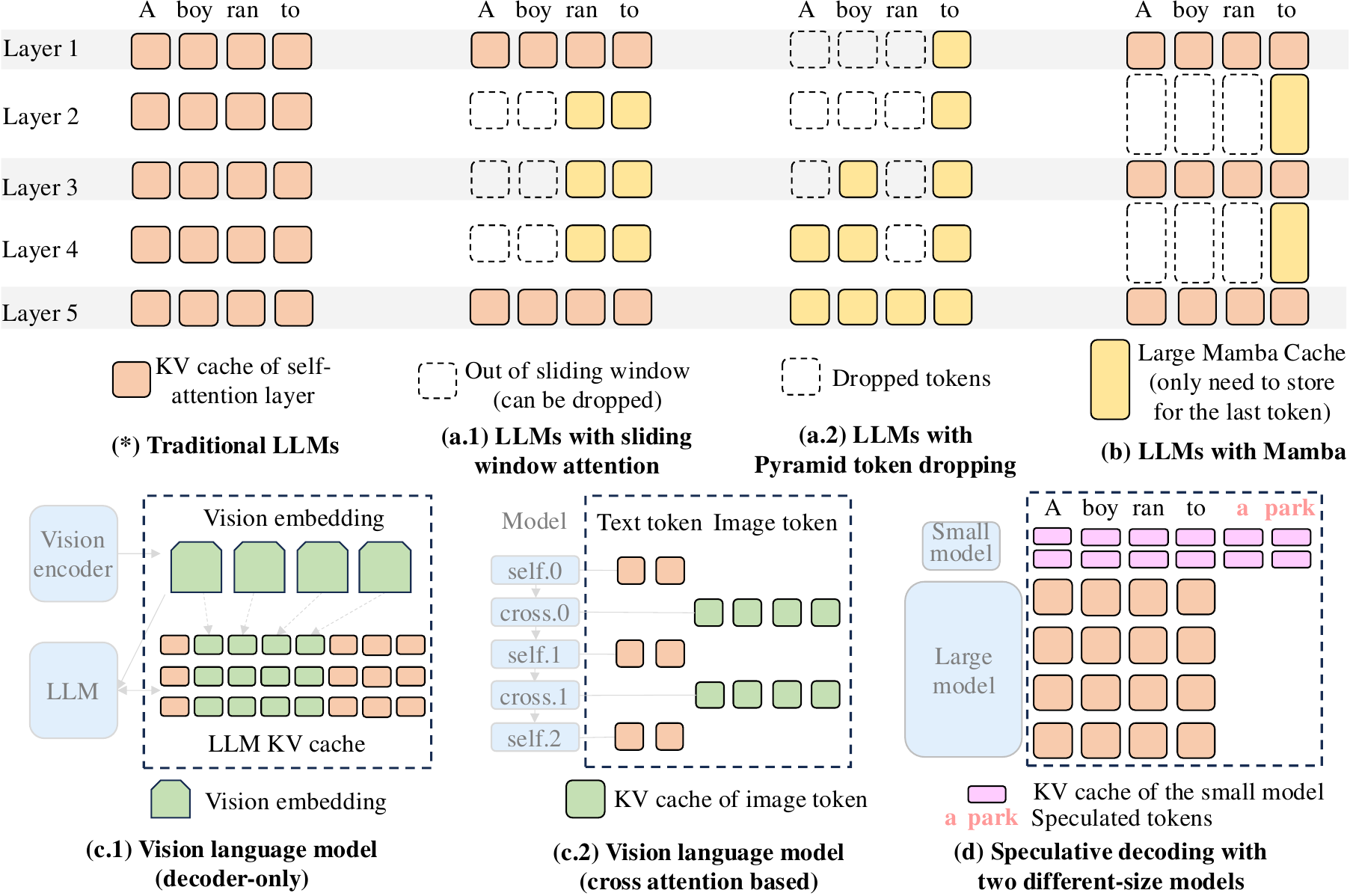}
    \caption{Contrasting traditional LLMs (top left) and latest LLMs. LLMs are becoming more and more heterogeneous: the KV cache sizes may differ, the KV cache dependencies are different, and the LLM architecture can also diverge}
    \label{fig:motivation}
\end{figure*}


\tightsection{Heterogeneous LLMs and Challenges}
\label{sec:motivation}

\vspace{0.1cm}

\tightsubsection{Heterogeneous LLMs}

Nowadays, the state-of-the-art LLMs go beyond stacking homogeneous full-context self-attention layers. Many new types of attention layers have been introduced to the LLMs, making LLM architecture heterogeneous. In this subsection, we discuss four sources of heterogeneity, as shown in \Cref{fig:motivation}.

\para{(a) Sparse attention} In the traditional self-attention layer, the KV cache size grows linearly with respect to the request length.
Sparse attention variants aim to reduce the KV cache size by attending to only a subset of prefix tokens.
The widely adopted version of sparse attention is \emph{sliding-window attention}, where each token only attends to a fixed number of adjacent tokens inside the sliding window.
To trade-off between model quality and KV cache size, recent models typically use a mix of full- and sliding-window attention layers (\Cref{fig:motivation}\cgreen{a.1}), including Google's \gemma~\cite{team2024gemma} and Mistral AI's Ministral model~\cite{ministral}. 
More advanced sparse attention variants, such as dynamically dropping some of the tokens~\cite{yang2024pyramidinfer,cai2024pyramidkv} (\Cref{fig:motivation}\cgreen{a.2}), are also proposed to control the KV cache size.

\para{(b) State space models}~\cite{gu2023mamba}, or linear attention models~\cite{katharopoulos2020transformers,orvieto2023resurrecting,peng2023rwkv}, take the idea of sparse attention to the extreme: for every token, it uses a large but fixed-size tensor to capture the context information of its previous tokens. These tensors are updated recurrently during decoding. 
These layers are also mixed with self-attention layers (as in Jamba~\cite{lieber2024jamba}, \Cref{fig:motivation}\cgreen{b}). 
Thus, the memory allocator needs to coordinate two different patterns, i.e., a small number of large fixed-size states, one for each state space layer, and a large number of small KV cache blocks, one for each token of each self-attention layer.

\para{(c) Multi-modal language models} typically accept inputs of multiple modalities in addition to text as input and generate text output. We show an example of such Vision Language Model (VLM) in \Cref{fig:motivation}\cgreen{c.1}~\cite{liu2024visual}. 
This VLM contains a \textit{vision encoder} that takes images as input and generates vision embeddings in the format of \emph{image tokens}. Then, the \textit{LLM} merges image tokens and text tokens into one sequence and performs autoregressive text generation with self-attention layers as normal LLMs. 
The memory allocator needs to manage the vision embedding cache, which only contains image tokens, and the KV cache of LLM parts, which contains both text tokens and image tokens. Due to KV cache compression techniques such as grouped query attention (GQA)~\cite{ainslie2023gqa}, the KV cache size of a token also differs from the size of the vision embedding of an image token. 

Moreover, as shown in \Cref{fig:motivation}\cgreen{c.2}, some VLMs, including Meta's \llama 3.2 vision model~\cite{dubey2024llama} and NVIDIA's NVLM model~\cite{dai2024nvlm}, utilize cross-attention to integrate the results from the image encoder into the text decoder. These LLMs have interleaving self-attention within text tokens (with KV cache for text tokens) and cross-attention between image tokens and text tokens (with encoder KV cache for image tokens). These two types of tokens can have different KV cache sizes.

\para{(d) Serving multiple concurrent models} There is also the need to serve multiple models inside a single LLM inference engine. An example is speculative decoding (\Cref{fig:motivation}\cgreen{d}). It \textcircled{\raisebox{-0.9pt}{1}} uses a small model to quickly propose new tokens sequentially, and \textcircled{\raisebox{-0.9pt}{2}} uses a large model to verify the correctness of the tokens in parallel so that it can generate multiple new tokens in each forwarding pass of the large model and keep the same quality. The KV cache size of each token differs a lot between the two models.

\tightsubsection{Heterogeneous KV Cache Size Causes Memory Fragmentation}
The above heterogeneity leads to the need to allocate a KV cache of different sizes for different tokens. In this section, we analyze the fragmentation of PagedAttention when serving heterogeneous LLMs using \llama 3.2 11B Vision model as an example. This model contains 32 self-attention layers, which require KV cache for text tokens, and eight cross-attention layers, which require KV cache for image tokens.





\begin{figure}
    \centering
    \includegraphics[width=0.45\textwidth]{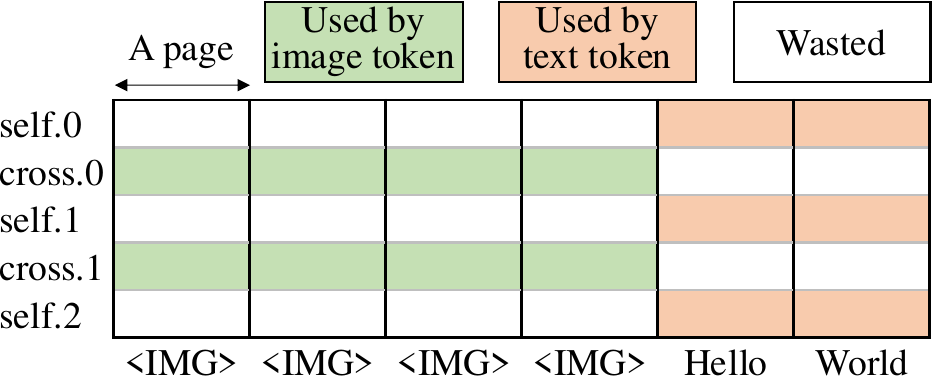}
    \caption{Visualizing the memory waste of \llama 3.2 vision model with  2 cross-attention layers (image tokens) and 3 self-attention layers (text tokens).}
    \label{fig:mllama_waste}
\end{figure}

Since the original PagedAttention can only deal with homogeneous layers, it needs to allocate KV cache for both text and image tokens for all layers (\Cref{fig:mllama_waste}). Suppose a request has $T$ text tokens and $I$ image tokens, the embedding size per layer per token is $E$ bytes, then PagedAttention needs to store
$
(T  + I) \times (32 + 8) \times E
$
bytes of memory, while ideally, we only need to store text token KV cache for self-attention layers and image token KV cache for cross-attention layers, and the necessary memory for a single request should be
$
(T \times 32 + I \times 8) \times E
$.

In the MMMU-pro~\cite{yue2024mmmu} dataset, which contains 6193 image tokens and 43  text tokens for each request on average, the resulting memory waste is 79.6\%. Similarly, the memory waste of \gemma and Ministral, two models that combine self-attention and sliding window attention, is up to 25\% and 56.25\%, respectively. Therefore, a new memory allocator is needed to reduce the memory waste of heterogeneous LLMs.







\tightsubsection{Heterogeneous Dependency Leads to Challenges in Prefix Caching}
\label{sec:challenge-prefix}

In addition to memory fragmentation, heterogeneous dependency patterns also introduces challenges for prefix caching. We summarize the challenges in prefix caching as follows:

\para{Different hit and eviction rules across layer types}  Different layer types have distinct cache hit rules due to their different token-dependency patterns. 
Self-attention layers compute attention between a token and all its prefixes, requiring all prefix tokens to remain unevicted to achieve a cache hit. In contrast, efficient attention mechanisms, such as those using sliding windows, only attend to a subset of prefix tokens to generate a new token. A cache hit occurs as long as this subset remains unevicted. For example, consider a prompt [\sout{token1}, token2, token3, \sout{token4}], where \sout{token} indicates an evicted token. When the sliding window size is 2, [token1, token2, token3] is still a valid prefix cache hit because token1 lies outside the sliding window, and its KV cache is not needed. 

These differences in cache hit rules also necessitate customized eviction policies. For example, in sliding window layers, tokens outside the window should be prioritized for eviction over the most recent tokens.

\begin{figure}
    \centering
    \includegraphics[width=0.45\textwidth]{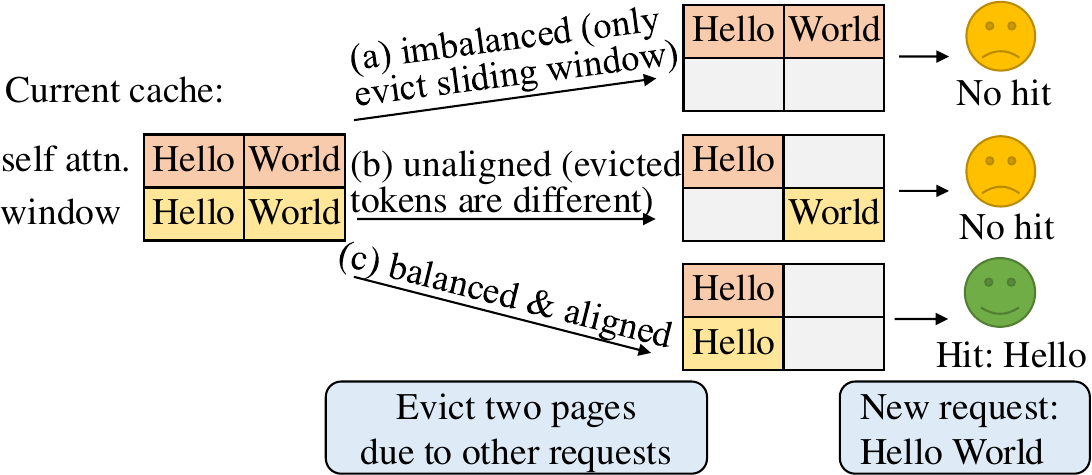}
    \vspace{-0.5em}
    \caption{Balanced and aligned cache eviction policy can improve hit rate.}
    \vspace{-0.5em}
    \label{fig:align_eviction}
\end{figure}

\para{Balanced eviction across different types} It is important to balance the number of evicted tokens across different layer types. A model-wise prefix cache hit requires the prefix to exist in the prefix cache of all layer types. If one layer type evicts too many tokens, such as the sliding window layer in \Cref{fig:align_eviction}\cgreen{a}, it can prevent a cache hit, even if the prefix remains in the KV caches of other layers. However, balance does not imply evicting the same number of tokens for all layers. Different layers need different numbers of tokens to achieve similar cache hit rates, and the eviction strategy should consider the unique properties of each layer type to optimize overall performance.



\para{Aligned eviction of different types} Eviction policies across layers must be aligned to ensure that similar sets of tokens are evicted. Each token is represented by multiple pages, one for each layer type, and a cache hit of that token requires it to remain unevicted in all types. If different layers evict different sets of tokens (``world'' of self-attention layer and ``hello'' of sliding window layer in \Cref{fig:align_eviction}\cgreen{b}), tokens inside the union of these sets will become unable to hit, and overall prefix cache hit rate is reduced. To address this, cache eviction policies need to be aligned across layer types, ensuring that similar sets of tokens are evicted to maximize the cache hit rate.

\tightsection{Two-level Memory Allocation}
\label{sec:allocation}

\vspace{0.1cm}

\tightsubsection{Overview}
\begin{figure}[t]
    \centering
    \includegraphics[width=0.45\textwidth]{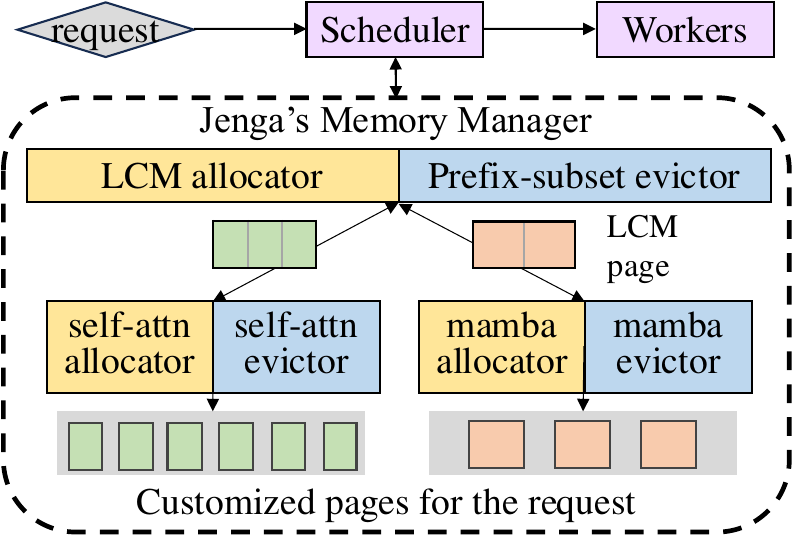}
    \vspace{-0.5em}
    \caption{Overview of \sys: a two-level memory management system for different types of layers. \sys is composed of the LCM allocator for first-level page allocation and the prefix subset evictor for page deallocation. Within the page, a customized allocator and evictor manage the memory for the specific layer type.}
    \label{fig:overview}
\end{figure}
\begin{figure}[t]
    \centering
    \includegraphics[width=0.45\textwidth]{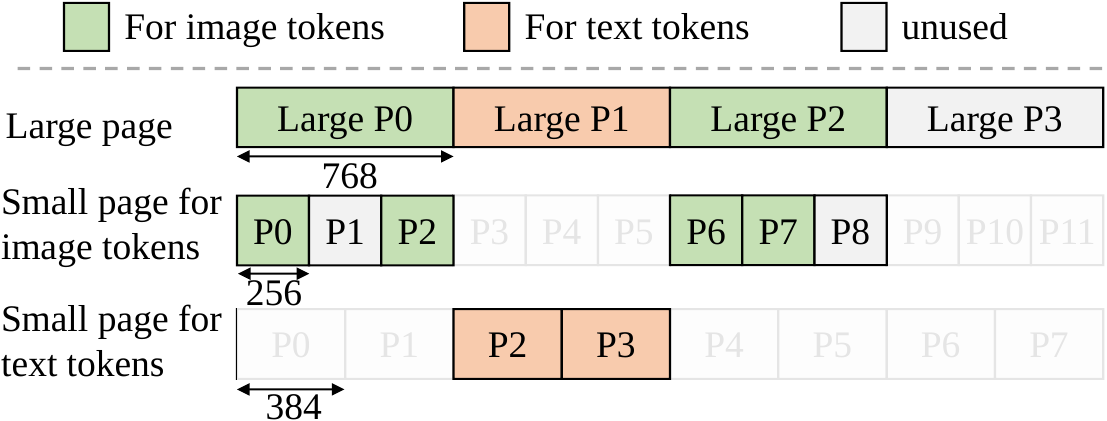}
    \vspace{-0.5em}
    \caption{Two-level allocation for \llama 3.2 vision model. }
    \label{fig:two_level_overview}
\end{figure}

The heterogeneity of LLMs, as discussed in \Cref{sec:motivation}, motivates \sys, a two-level memory management system that allocates memory for different types of layers by introducing a compatibility layer and a customization layer. The overview of \sys is shown in \Cref{fig:overview}.

For memory allocation, \sys introduces the \textit{LCM allocator} to allocate pages with sizes compatible across all layer types, and \textit{customized allocators} for each specific layer type (e.g., self-attention and mamba). The customized allocators allocate pages with the specialized page size of their type from the compatible pages. For prefix cache management, \sys introduces a \textit{prefix-subset evictor} to coordinate the eviction among different layer types, and \textit{customized evictors} to customize the eviction strategy of each type.

\Cref{fig:two_level_overview} shows the memory layout of \sys for \llama vision model with page size 256 for image tokens and 384 for text tokens.\footnote{We assume the KV cache size per token of each layer is 128, and the model contains 2 cross attention layers (with KV size per image token $128\times 2=256$) plus 3 self-attention layers (with KV size per text token $128 \times 3 = 384$). For simplicity of explanation, we set $tokens\_per\_page=1$ in this paper, but \sys is capable of arbitrary $tokens\_per\_page$.} \sys uses the LCM of all page sizes as the compatible page size, which is $LCM(256,384)=768$ in this case. We will compare LCM with other potential options of the compatible page size, e.g., the GCD or MAX of all layers, in \Cref{sec:choices}. \sys first partitions the entire KV Cache memory into large pages of LCM size and uses the LCM allocator to manage them. The customized allocators request some large pages from the LCM allocator (large pages 0 and 2 for image tokens, large page 1 for text tokens), partition the large pages into small pages tailored to that type, and allocate the small pages as needed (the 4 256-byte small pages for 4 image tokens and 2 384-byte small pages for 2 text tokens in request \texttt{<IMG><IMG><IMG><IMG>Hello World}).

Specifically, the customized small page allocators interact with the LCM allocator as follows: 

\begin{itemize}
    \item \texttt{allocate()} for allocating a small page of that type. If all small pages for this customized allocator are allocated, it requests a new large page from the LCM allocator and partitions it into free small pages. Then, the allocator allocates an unused small page.
    \item \texttt{free(small\_page\_id)} to free a small page. The customized allocator marks the small page as unused. If all small pages within a large page are unused, the large page is returned to the LCM allocator.
\end{itemize}

The LCM-based two-level allocation strategy prevents external fragmentation among large pages, and \Cref{sec:req-aware} can reduce internal fragmentation inside each large page. Additionally, for each layer type, allocated pages can be fully represented by small page IDs of that type (e.g. small pages P2 and P3 for text tokens) so the attention kernels can be executed as if there is only one layer type and do not need to consider the complexity introduced by multiple types.

\tightsubsection{Execution with New Memory Layout}
\label{sec:new-layout}
Although \sys employs a memory layout distinct from the standard PagedAttention, \sys can reuse the PagedAttention workers with very little change. This section explains how \sys works with existing PagedAttention workers.
\begin{figure}[t]
\begin{subfigure}{.5\textwidth}
    \centering
    \includegraphics[width=.9\linewidth]{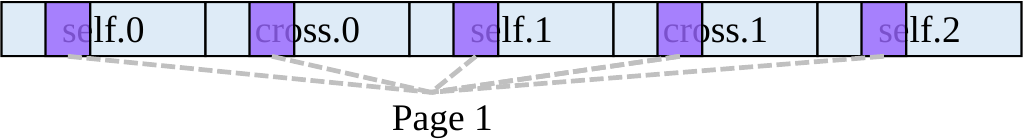}
    \vspace{-0.5em}
    \caption{Memory layout of PagedAttention}
    \label{fig:logic_page}
\end{subfigure}
\begin{subfigure}{.5\textwidth}
    \centering
    \includegraphics[width=.9\linewidth]{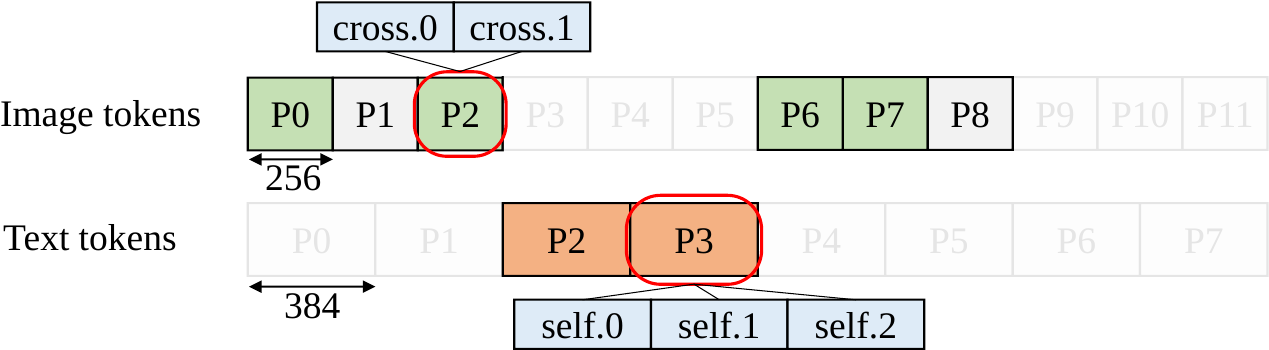}
    \vspace{-0.5em}
    \caption{Memory layout of \sys}
    \label{fig:exec_layout}
\end{subfigure}
\begin{subfigure}{.5\textwidth}
    \centering
    \includegraphics[width=.9\linewidth]{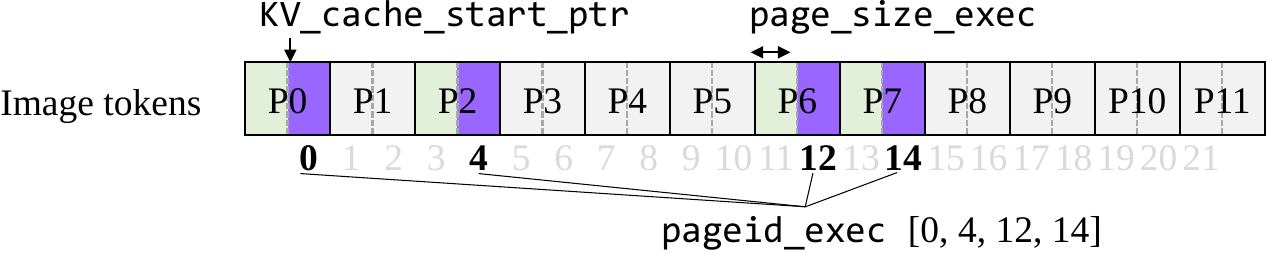}
    \vspace{-0.5em}
    \caption{Memory allocated for layer \texttt{cross.1}}
    \label{fig:exec_cross_1}
\end{subfigure}
\vspace{-2em}
\caption{Memory layout of PagedAttention and \sys}
\label{fig:traverse}
\end{figure}

\para{New memory layout for inter-type memory exchange} As shown in \Cref{fig:logic_page}, in standard PagedAttention, each logical page is divided into multiple slices in the physical memory, making it difficult to exchange pages between different types of KV cache. Specifically, the memory layout of the standard PagedAttention follows a \textit{layer-page partition} that first partitions the memory into layers and then partitions each layer into pages. This layout simplifies model execution because when executing one layer, we only need to pass the memory of that layer. This layout is widely used by inference engines including vLLM~\cite{kwon2023efficient}, SGLang~\cite{zheng2023efficiently}, TGI~\cite{tgi}, and attention libraries, e.g., FlashAttention~\cite{dao2022flashattention}, FlashInfer~\cite{flashinfer}, but does not meet the need of \sys.

To enable memory sharing between memory types, \sys proposes a \textit{page-layer partition} to make each small page consecutive. As shown in \Cref{fig:exec_layout}, \sys partitions the memory into pages and then partitions each page into layers.

Despite this new memory layout, \sys is still compatible with PagedAttention workers and kernels. As shown in \Cref{fig:exec_cross_1}, memory allocated for a layer (e.g., \texttt{cross.1}) can still be represented by a customized \texttt{start\_ptr}, \texttt{page\_size}, and \texttt{page\_id}, maintaining consistency with the PagedAttention kernel interface.

\para{\sys using PagedAttention workers with minimal change} \sys requires no modifications to PagedAttention kernels. Attention can be computed as usual by passing the above \texttt{kv\_cache\_start\_ptr}, \texttt{page\_size\_exec}, \texttt{pageid\_exec} to the libraries.

The changes to the inference engine workers are also minimal. Major changes only include: (1) allocate a single KV cache tensor and assign different offsets (\texttt{KV\_cache\_start\_ptr}) for each layer, instead of allocating fixed-size tensors for each layer (2) prepare attention metadata (parameters passed to attention kernels, e.g., page ID), for each layer type, rather than a global metadata for all layers.

\tightsubsection{Request Aware Allocation to Reduce Internal Fragmentation of Large Pages}
\label{sec:req-aware}
\begin{figure}[t]
    \centering
    \includegraphics[width=0.46\textwidth]{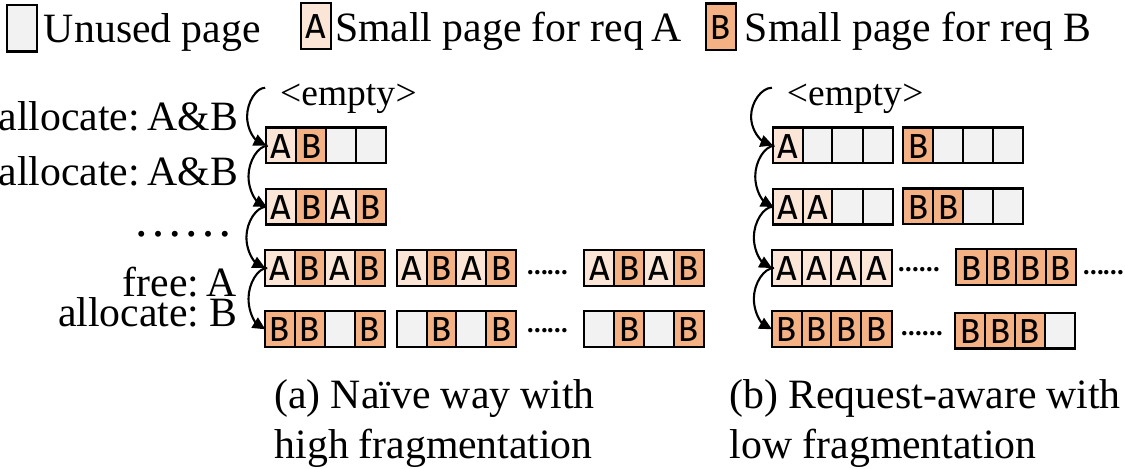}
    \vspace{-1em}
    \caption{Internal fragmentation with two requests A\&B, and 4 small pages per large page}
    \label{fig:internal_frag}
\end{figure}

A naive allocation strategy can lead to large internal fragmentation due to the different allocation and free patterns in LLM workloads. As shown in \Cref{fig:internal_frag}\cgreen{a}, memory allocation for a single request is often interleaved with allocations for other requests, whereas memory deallocation for a single request typically occurs together.  When a request is completed, numerous small pages will be freed. However, only very few large pages can be returned to the large-page allocator because the small pages of that request are sharing the large pages with other requests due to the interleaved allocation. This results in severe internal fragmentation.

\sys addresses the memory fragmentation problem by aligning the allocation and deallocation patterns for each request. Specifically,  as shown in \Cref{fig:internal_frag}\cgreen{b}, \sys allocates small pages within a single large page to the same request whenever feasible. This ensures that large pages can be returned to the large-page allocator once the request is completed. For attention variants where the freeing of different pages within the same request does not occur simultaneously, this approach remains effective, as adjacent small pages are typically freed shortly after one another.


The complete allocation algorithm considering the interaction between the allocator for the two levels, as well as the request aware allocation, is shown below. Each large page and all its small pages are associated with a specific request. \sys prioritizes allocating small pages to their associated requests by the following algorithm:
\begin{enumerate}
    \item Allocate an unused small page associated with that request.
    \item If fail, request a new large page from the large-page allocator, mark all small pages inside the large page as associated with this request, and allocate one of these small pages.
    \item If steps 1-2 fail, allocate an unused small page in this small-page allocator but associated with other requests.
\end{enumerate}

\tightsubsection{Discussion: Different Choices of Compatibility Layer}
\label{sec:choices}
The aligned embedding size can motivate many different design decisions of the compatibility layer, each with its own advantages and limitations. We think that LCM is the most flexible and extensible choice among them.

\para{GCD page} Using the greatest common denominator (GCD) of different embedding sizes as the compatible page size. This approach will have no internal fragmentation. However, it significantly reduces LLM inference speed. This is because the most efficient GPU kernels typically require the KV cache to be contiguous along specific tensor dimensions. The GCD solution may have to partition the tensor along these dimensions, requiring the customization of GPU kernels for a wide range of GCDs. This greatly increases GPU kernel engineering overhead, and even with such customization, performance often falls short of that achieved by the most efficient kernels. For example, MuxServe~\cite{duan2024muxserve} uses a GCD page to serve multiple models but restricts itself to models with the same size per head to avoid excessive GPU kernel development.



\para{MAX page} Take the maximum of different embedding sizes as the compatible page size. This solution will have internal fragmentation for layers with smaller page size. A potential workaround is to increase the number of tokens per page for these small layers. However, this results in coarser granularity for both memory allocation and cache hits. For example, Jamba 52B's large mamba state requires assigning 1344 tokens to each self-attention page to avoid internal fragmentation, which exceeds the typical number of tokens in real-world requests, such as 1085.04 on average in ShareGPT~\cite{sharegpt}.

\para{LCM page} Take the least common multiple (LCM) of different embedding sizes as the compatible page size, which is used by \sys. The LCM page does not need new GPU kernels or an extremely large number of tokens assigned to each page. However, it may lead to internal fragmentation within each large page due to unused small pages. 
\sys addresses this problem by request-aware allocation (\Cref{sec:req-aware}). Another potential problem is that the LCM might be extremely large. In practice, for all models supported by \latestvllm, the largest LCM comes from Jamba, where it is $84\times$ the small page size used in the model, and we do not observe any performance degradation in that model. 

\begin{figure}[t]
\small
\begin{subfigure}{.47\textwidth}
\centering
\begin{minted}[linenos=false]{python}
class LayerSupportsPrefixCache:
  def update_last_access(r: Request, time: int);
  def set_prefix_length(r: Request);
  def get_possible_prefix(is_hit: List[bool]);
\end{minted}
\vspace{-1em}
\caption{Interface for customized prefix caching of different layer type}
\label{fig:interface/main}
\end{subfigure}
\begin{subfigure}{.47\textwidth}
\centering
\begin{minted}[linenos=false,mathescape]{python}
class SlidingWindowLayer(LayerSupportsPrefixCache):
  def update_last_access(r: Request, time: int):
    for i in range(r.len-sliding+1, r.len+1):
      self.evictor.update_last_access(r.page[i], time)

  def set_prefix_length(r: Request):
    for i in range(0, r.len):
      self.evictor.set_prefix_length(r.page[i], i)

  def get_possible_prefix(is_hit: List[bool]):
    return @$\{ p \mid \forall x \in [0, \texttt{sliding}), \; \texttt{is\_hit[p-x] is True} \}$@
\end{minted}
\vspace{-1em}
\caption{Prefix caching support of sliding window layers}
\label{fig:interface/sliding}
\end{subfigure}
\vspace{-1em}
\caption{Layer property aware prefix caching}
\label{fig:interface}
\end{figure}
\tightsection{Customizable Prefix Caching}
\label{sec:prefix-cache}

The prefix caching system of an inference engine needs to support two tasks:
\begin{enumerate}
    \item \textbf{Cache eviction:} Evict an existing page from the cache to free up space for a new page.
    \item \textbf{Cache hit:} Identifying the cached prefix for a request
\end{enumerate} 

For both tasks, the expected behavior varies across different layer types, and thus needed to be customized inside the inference engine. \sys provides unified interfaces to customize each layer, and a global prefix-subset evictor that manages the diverse layer types by invoking these APIs.
The interface is shown in \Cref{fig:interface/main}, with \texttt{update\_last\_access} and \texttt{set\_prefix\_length} for customized eviction rule, and \texttt{get\_possible\_prefix} for customized hit rule. Further details on cache eviction and cache hit mechanisms are provided in \Cref{sec:cache-evict} and \Cref{sec:cache-hit}.

\tightsubsection{Customizable Cache Eviction}
\label{sec:cache-evict}

As discussed in \Cref{sec:challenge-prefix},  \sys's cache eviction policy must ensure both balance and alignment across different layer types. This section illustrates \sys's approach using least recently used (LRU) eviction as an example. 

\para{Balanced eviction by \texttt{update\_last\_access}} \sys provides a coarse grain API, \texttt{update\_last\_access}, to set unified last-access times across different layers, thus enabling balanced eviction. In LRU eviction, the page with the earliest last-access time is evicted. By aligning last-access times for tokens in the same request but different layers, \sys ensures that eviction priorities across layers remain similar.

\begin{figure}[t]
\centering
\begin{subfigure}{.46\textwidth}
\centering
\resizebox{0.9\linewidth}{!}{
\begin{tabular}{cccc}
    \toprule
       Step  & Request  & self attention & sliding window\\ \midrule
         1 & 1's prefill  & ABCD->E & ABCD->E\\
         2 & 1's decode  & ABCDE->F & DE->F\\ \midrule
         3 & 2's prefill  & ABCDG->H & CDG->H \\ \bottomrule
         
    \end{tabular}}
    \caption{Timeline of the 2 requests. ABC->D means access KV cache of tokens ABC and generate token D. Note that the generated token does not have KV cache. Assume sliding window size 2.}
    \label{tab:my_label}
\end{subfigure}
\begin{subfigure}{.22\textwidth}
    \centering
    \includegraphics[width=.9\linewidth]{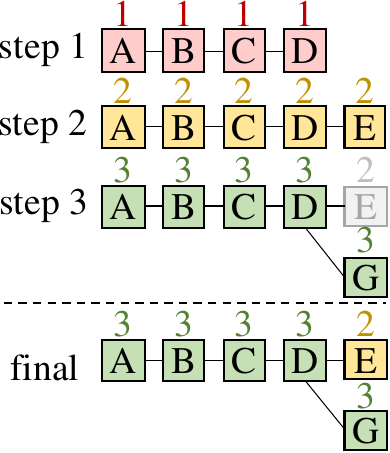}
    \vspace{-0.5em}
    \caption{Self-attention layer}
    \label{fig:last-access-req1}
\end{subfigure}
\begin{subfigure}{.22\textwidth}
    \centering
    \includegraphics[width=.9\linewidth]{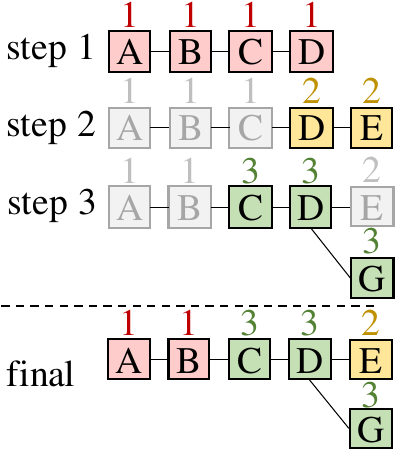}
    \vspace{-0.5em}
    \caption{Sliding window layer}
    \label{fig:last-access-req2}
\end{subfigure}
\vspace{-0.5em}
\caption{Last access time after two requests}
\label{fig:last-access}
\end{figure}

\Cref{fig:last-access} shows the timeline and last access time of a model with one self-attention layer and one sliding window layer when running the following two requests:

\begin{itemize}
    \item Request 1: input \texttt{[\cgreen{A B C D}]} and output \texttt{[E F]}
    \item Request 2: input \texttt{[\cgreen{A B C D} G]} and output \texttt{[H]}
\end{itemize}

For simplicity of explanation, we assume all layer types have the same page size, such that each ``large page'' contains one ``small page''. In this scenario, the terms ``large page'' and ``small page'' can be used interchangeably and are collectively referred to as ``page.'' The general case, where multiple page sizes coexist, is addressed in \Cref{sec:evict-lcm}.

For self-attention layer, the last access times of all running tokens are updated in each step. For example, in \Cref{fig:last-access-req1}, tokens [A B C D] are updated in step 1, and then tokens [A B C D E] in step 2 and [A B C D G] in step 3.

For the sliding-window layer, \sys updates the last access time only for tokens within the sliding window, as these are the tokens actively involved in the attention computation. For example, in step 2, tokens [A B C] are not updated because generating token [F] only needs the KV cache of tokens [D E]. This customization ensures that tokens outside the sliding window retain an earlier last access time, making them a higher priority for eviction. 

Despite these customizations, eviction remains balanced since tokens from the same request share identical last-access timestamps across layers. For example, \sys will evict tokens exclusive to Request 1 (e.g., [E] with time\_stamp 2) in both layers before those from Request 2 (e.g., [C D G] with time\_stamp 3) based on the last access time.


\para{Aligned eviction by \texttt{set\_prefix\_length}} \sys provides a fine-grained API, \texttt{set\_prefix\_length}, to refine eviction priorities for pages with the same last-access time to ensure aligned eviction. By assigning identical prefix length values to the page of the corresponding token across layers, \sys ensures consistent eviction priorities of these pages. For example, tokens [A B C D G] in both layers can be assigned lengths [1, 2, 3, 4, 5], respectively. Thus, the token with the highest length (e.g., [G] in the two layers) is evicted before other tokens (e.g., [C D]),  maintaining alignment across layers.



\tightsubsection{Customizable Cache Hit}
\label{sec:cache-hit}
Cache hit rules differ across layer types. For example, a prefix hit in a self-attention layer requires all tokens in the prefix to remain unevicted, whereas a sliding window layer only requires the last sliding\_window\_size tokens of the prefix to remain unevicted.

\begin{figure}[t]
\begin{subfigure}{.5\textwidth}
    \centering
    \includegraphics[width=.8\linewidth]{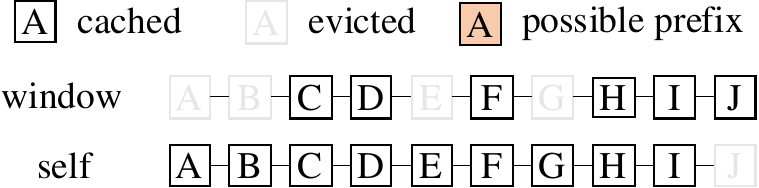}
    \caption{The KV cache}
    \label{fig:hitcontrol-cache}
\end{subfigure}
\begin{subfigure}{.5\textwidth}
    \centering
    \includegraphics[width=.8\linewidth]{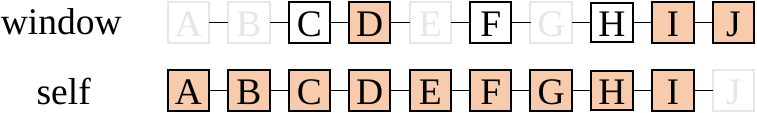}
    \caption{The possible prefixes for request \texttt{ABCDEFGHIJ}}
    \label{fig:hitcontrol-hit}
\end{subfigure}
\vspace{-1em}
\caption{Customizable cache hit} 
\label{fig:hitcontrol}
\end{figure}

\sys provides the \texttt{get\_possible\_prefix} API to custom the cache hit rules for different layer types. Given the parameter \texttt{is\_hit} indicating whether the KV value of each token is cached, the function should return all valid prefixes of that layer. For example, for request [ABCDEFGHIJ], with the KV cache status in \Cref{fig:hitcontrol-cache}, valid prefixes for a self-attention layer are [A], [AB], ..., [ABCDEFGHI],  while valid prefixes for a sliding window layer are [ABCD], [ABCDEFGHI], and [ABCDEFGHIJ]. 
The prefix [ABC] is invalid for a sliding window layer because both B and C must remain cached to achieve such hit.


Upon receiving a new request, the compatibility layer invokes \texttt{get\_possible\_prefix} for each layer type to identify valid prefixes. The longest common prefix valid across all layers is selected as the cache hit prefix.

\tightsubsection{The Customization of Different Layers}

\para{Sliding Window Layer} \sys updates the last-access time only for tokens within the sliding window and require only these tokens to remain cached for a prefix hit. Its implementation is shown in \Cref{fig:interface/sliding}.

\para{Mamba Layer} For Mamba layers, caching is specialized due to the significantly larger per-token state compared to other attention-based layers. Instead of caching all tokens, which requires too much memory, \sys only caches the state of every 512 tokens.
Then, \sys can hit the prefixes with length a multiplier of 512, which is implemented by the \texttt{get\_possible\_prefix} interface. Additionally, only the last cached token's access time is updated via \texttt{update\_last\_access}. We also notice Marconi~\cite{pan2024marconi}, a concurrent work that provides an advanced algorithm to select the set of tokens to cache. Its algorithm can be integrated into \sys to enhance the prefix caching.

\para{Vision Embedding Cache and Vision Cross Attention Cache} In these caches, evicting even a single image token will require the recomputation of the entire vision encoder. To minimize the number of recomputed images, it is better to evict all tokens from one image than to evict an equivalent number of tokens across multiple images. To achieve this, \sys assigns a randomized \texttt{prefix\_length} to each image and applies this value to all corresponding tokens via \texttt{set\_prefix\_length}. Tokens from the image with the highest random value are prioritized for eviction.

\tightsubsection{Cache Eviction of LCM Page Table}
\label{sec:evict-lcm}

\sys manages memory with a two-level approach: \textit{large pages} that are compatible across all layer types and \textit{small pages} that are customizable for specific layer types. This design enables \sys to achieve two benefits simultaneously:
\begin{enumerate}
    \item Fine-grained, customizable cache hits for small pages.
    \item Efficient memory sharing via coarse-grained large-pages.
\end{enumerate}
However, effective prefix caching requires careful coordination between fine-grained small-page cache hits and coarse-grained large-page exchanges between different layer types.

Specifically, each small page can be in one of three states: (1) \textit{empty} page with no valid KV cache and is not used by any requests; (2) \textit{evictable} page with valid KV cache and is not used by any requests (3) \textit{used} page that is used by running requests and thus unevictable. Moreover, we call a large page \textit{empty} if all its small pages are empty, and \textit{evictable} if all its small pages are evictable.

The allocation algorithm in \sys should (1) prioritize unused pages and perform eviction only when necessary (2) try its best to keep small pages inside a large page in the same state to avoid internal fragmentation like \Cref{fig:internal_frag}\cgreen{a}.

To allocate a new small page of a specific type for a particular request, \sys implements the following steps:
\begin{enumerate}
    \item \textbf{Allocate a request-associated unused small page.} Each small page is associated with a specific request (\Cref{sec:req-aware}). \sys first attempts to allocate an empty small page of the required type that is associated with the current request.
    \item \textbf{Allocate from an empty large page.} If no suitable small page is available, \sys requests an empty large page from the large-page allocator, associates its small pages with the request, and allocates one of these small pages.
    \item \textbf{Allocate by evicting a large page.} If no unused large page exists, \sys evicts an evictable large page using the LRU eviction policy. The LRU timestamp of a large page is set to the latest last access time among all its small pages. After eviction, all small pages within the large page are marked empty, and one of them is allocated.
    \item  \textbf{Allocate an arbitrary unused small page.} If steps 1–3 fail, \sys allocates an unused small page of the required type that may not be associated with the specific request.
    \item \textbf{Allocate by evicting a small page.} As a last resort, \sys evicts an evictable small page of the required type using the LRU eviction policy and allocates it.
\end{enumerate}

\tightsection{Case Studies}
\vspace{1em}
\tightsubsection{Speculative Decoding and Multiple Model}
Speculative decoding involves an extra small model (the speculator) in the LLM inference engine.
\sys can be naturally extended to support such case as it can allocate two different KV cache sizes for the two models automatically with negligible fragmentation.

Moreover, \sys can be extended to serve multiple models inside the same LLM inference engine. 
After registering all models with the \texttt{custom\_kv\_cache} API, \sys can have a compatible page size for all models, which can be the granularity to exchange pages between inference engines. We leave the full support of serving multiple models as a future work.

\tightsubsection{Vision Embedding Cache of VLMs}
\label{sec:vlm-cache}
As the vision embedding cache can be treated as another type of layer with a specific hidden size, \sys can automatically handle the different page sizes of vision embedding tokens and LLM KV cache tokens. 

Moreover, with \sys's customizable memory management, \textit{\textbf{vision embedding cache does not increase the peak memory consumption of the model}} when the vision embedding cache per token is smaller than the LLM KV cache size per token, which is held by all VLMs in \latestvllm.
After being generated by the vision encoder, the vision embedding cache will be consumed by chunked prefill steps of the LLM part. Current inference engines support chunked prefill of LLM part in two ways, in both of which \sys can effectively manage the vision embedding:

\begin{figure}[t]
\begin{subfigure}{.5\textwidth}
    \centering
    \includegraphics[width=.9\linewidth]{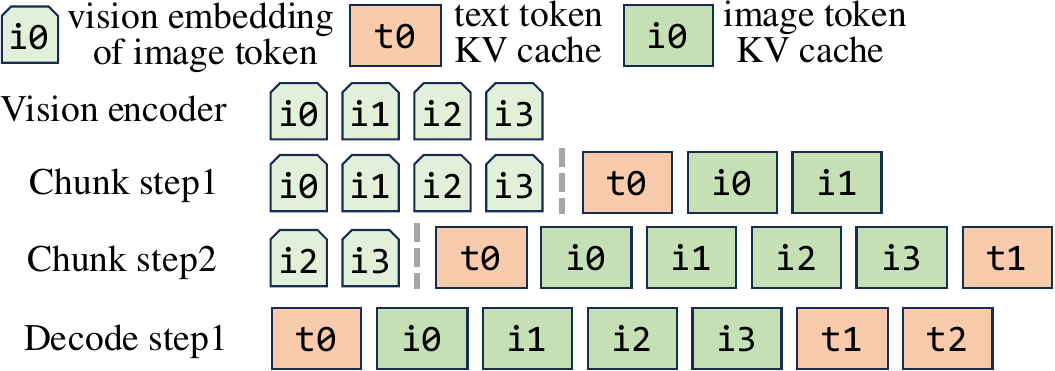}
    \caption{Vision embedding cache for allocate-on-demand KV cache}
    \label{fig:vision-ondemand}
\end{subfigure}
\begin{subfigure}{.5\textwidth}
    \centering
    \includegraphics[width=.65\linewidth]{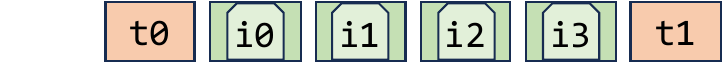}
    \caption{Vision embedding cache for fully-allocated KV cache}
    \label{fig:vision-reuse}
\end{subfigure}
\vspace{-2em}
\caption{Vision embedding and chunked prefill KV cache} 
\label{fig:hitcontrol}
\end{figure}

\para{Allocate-on-demand KV cache with free-on-demand vision embedding cache} Some inference engines only allocate the KV cache for tokens used in the current chunked prefill step. \sys can free the vision embedding of image tokens once consumed by the prefill step. Therefore, the peak memory usage is affected by the shrink of vision embedding and enlargement of the KV cache simultaneously and will not grow too large. \Cref{fig:vision-ondemand} shows the timeline of request \texttt{[t0 i0 i1 i2 i3 t1]}, where \texttt{t*} and \texttt{i*} refers to text token and image token respectively. \sys process this request with the following steps: (1) runs the vision encoder and generates the vision embedding; (2) runs the chunked prefill of 3 tokens \texttt{[t0 i0 i1]} and generates their KV cache; (3) frees the consumed vision embedding of image tokens \texttt{[i0 i1]} and runs the chunked prefill of 3 tokens \texttt{[i2 i3 t1]}; (4) frees the consumed vision embedding of image tokens \texttt{[i2 i3]} and runs a decode step. The peak memory usage is the KV cache of all tokens, plus at most \textit{chunk\_prefill\_size} tokens of vision embedding. The \textit{chunk\_prefill\_size} is much smaller than the number of tokens in all scheduled requests, and the vision embedding size per token is small compared to the KV cache size per token, so \sys can almost remove the memory consumption of vision embedding cache.

\para{Fully allocated KV cache and cache reusing by vision embedding} Other inference engines allocate the full KV cache of all prefill tokens at the first chunked prefill step. As the vision embedding cache of one token is used before generating the KV cache of that token, \sys supports reusing the KV cache for vision embedding cache, as shown in \Cref{fig:vision-reuse}, which completely removes the memory consumption of vision embedding cache.



\tightsection{Evaluation}
\label{sec:eval}


\sys is implemented with about 4000 lines of Python code on top of vLLM and does not require any change of CUDA kernels. \sys is compatible with all the 90 models in \latestvllm. \sys is transparent to users of the inference engine. \sys can parse all possible embedding sizes from the model structure and perform the memory allocation automatically. In this section, we evaluate the performance of \sys.

\tightsubsection{Evaluation Setup}

\textbf{Platform} We evaluate \sys on two GPU platforms: (1) NVIDIA H100 80GB GPU with 2 Intel Xeon Platinum 8480C CPUs, equipped with CUDA 12.5. This is the default platform in the evaluation unless explicitly mentioned (2) NVIDIA L4 24GB GPU with 2 AMD EPYC 7F52 16-Core Processor, equipped with CUDA 12.4.
\begin{figure*}[t]

\centering
\begin{subfigure}{.45\textwidth}
    \centering
    \includegraphics[width=\linewidth]{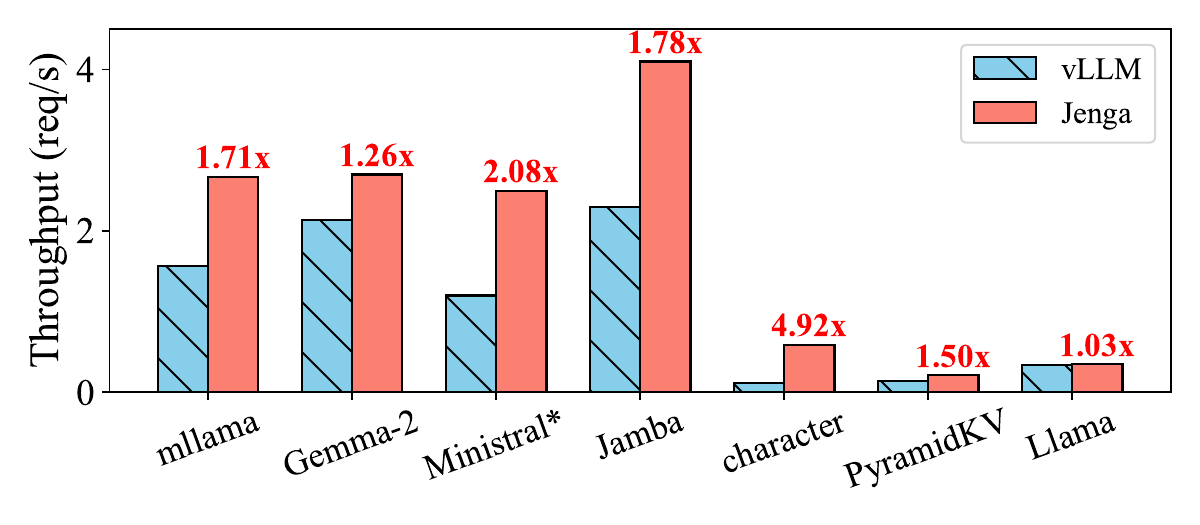}
    \vspace{-2.5em}
    \caption{H100 GPU}
    \label{fig:eva-e2e-throughput-h100}
\end{subfigure}
\hspace{2em}
\begin{subfigure}{.45\textwidth}
    \centering
    \includegraphics[width=\linewidth]{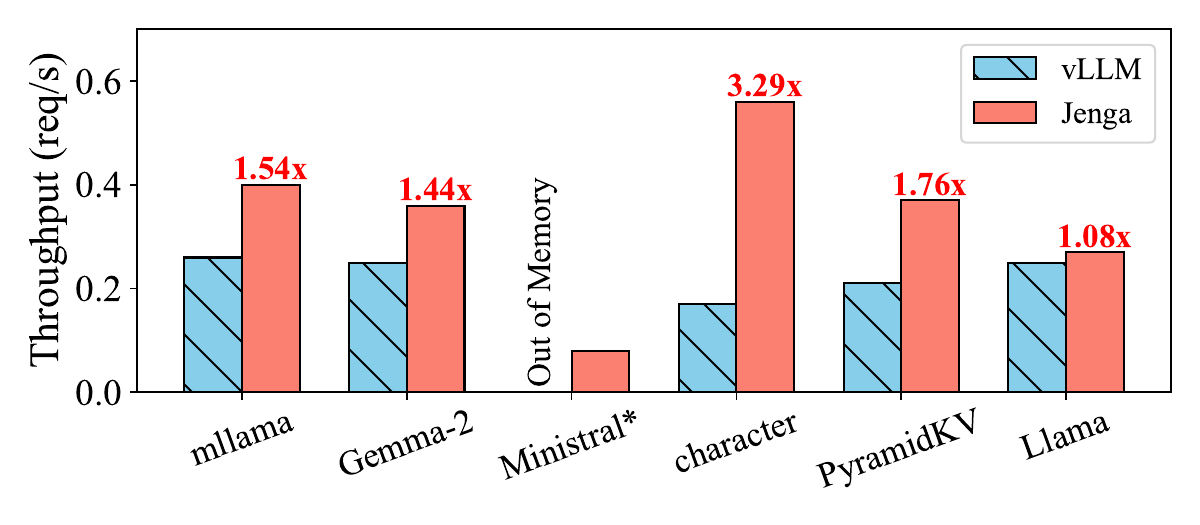}
    \vspace{-2.5em}
    \caption{L4 GPU}
    \label{fig:eva-e2e-throughput-l4}
\end{subfigure}
\vspace{-0.8em}
\caption{End-to-end throughput. Amplified Ministral's throughput in both vLLM and \sys by $10\times$ for better visualization.}
\label{fig:eva-e2e-throughput}
\end{figure*}
\begin{figure*}[t]
\begin{subfigure}{.33\textwidth}
    \centering
    \includegraphics[width=\linewidth]{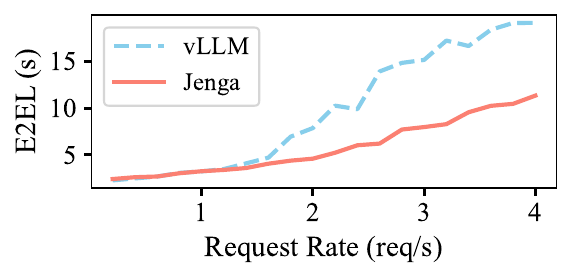}
    \label{fig:eva-e2e-e2el}
\end{subfigure}
\begin{subfigure}{.33\textwidth}
    \centering
    \includegraphics[width=\linewidth]{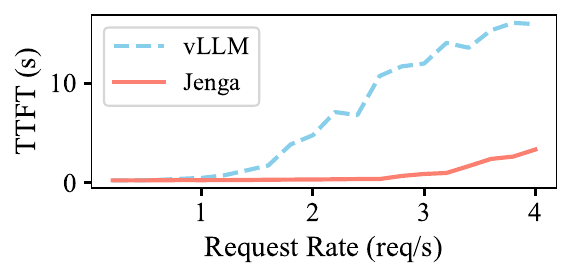}
    \label{fig:eva-e2e-ttft}
\end{subfigure}
\begin{subfigure}{.33\textwidth}
    \centering
    \includegraphics[width=\linewidth]{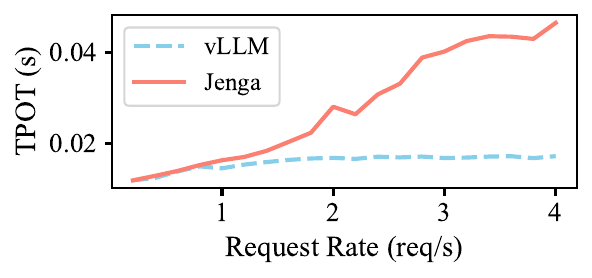}
    \label{fig:eva-e2e-tpot}
\end{subfigure}
\vspace{-3em}
\caption{Averaged Latency for the Llama Vision Model (mllama) with changing request rates. E2EL denotes end-to-end latency, TTFT denotes time to first token, and TPOT denotes time per output token.}
\vspace{-1em}
\label{fig:eva-e2e-latency}
\end{figure*}
\begin{table}[]
    \centering
    \footnotesize
    \begin{tabular}{ccccc} \toprule
         Model &	 Dataset & H100 & L4 \\ \midrule
\llama 3.2 Vision (mllama)  & MMMU-pro & 11B & 11B$^{*}$ \\
\gemma &  arXiv-QA & 27B & 9B  \\
Ministral &   arXiv-QA & 8B & 8B$^{*}$ \\
Jamba-1.5 &  MMLU-pro  & 52B$^{*}$  & OOM \\
character.ai  & MMLU-pro & 70B$^{*}$ & 8B \\
PyramidKV & MMLU-pro & 70B$^{*}$ & 8B\\
\llama 3.1  & MMLU-pro & 70B$^{*}$ & 8B \\ \bottomrule
    \end{tabular}
    \vspace{-1em}
    \caption{Model and dataset. $^{*}$ means with FP8 quantization.}
    \label{tab:eva-setup}
\end{table}

\noindent\textbf{Baselines} We perform end-to-end evaluation of \sys by using vLLM v0.6.3 and only change the memory management system. We also offers a break down experiment to compare \sys with the memory management system of other state-of-the-art LLM inference engines, including vLLM~\cite{kwon2023efficient}, SGLang~\cite{zheng2023efficiently}, and TGI~\cite{tgi}. We don't perform end-to-end evaluation on these engines as they only support a very small subset of the evaluated models.

\noindent\textbf{Models} We include a wide range of heterogenous LLMs. \llama vision model~\cite{dubey2024llama} is a multi-modal model with cross attention layers. \gemma~\cite{team2024gemma} and Ministral~\cite{ministral} are two models with sliding window layers. Jamba~\cite{lieber2024jamba} contains Mamba~\cite{gu2023mamba} layers. Character.ai~\cite{character} is a private model with sliding window layers and KV cache sharing. We implement the model based on their blog post on top of \llama. PyramidKV~\cite{yang2024pyramidinfer} is a sparse attention model that drops different number of tokens in different layers. All these models mix the aforementioned attention variants with standard self-attention layers. We also use standard \llama with self attention layers only for evaluating the overhead of \sys. 
The size of the model is listed in \Cref{tab:eva-setup}. We also use LLaVA-OneVision~\cite{li2024llava} 7B, InternVL2~\cite{chen2024internvl} 8B, Phi-3 Vision~\cite{abdin2024phi} 4B, and Paligemma2~\cite{steiner2024paligemma} 10B to evaluate the vision embedding cache. Note that Paligemma2 is a model mixed with three types of memory, e.g.,  vision embedding cache, sliding window KV cache and self-attention KV cache. We do not evaluate Hamba~\cite{dong2024hymba} which is discussed in \Cref{sec:intro} because it lacks essential GPU kernels and is not supported by vLLM yet.

\noindent\textbf{Dataset} We use MMLU-pro~\cite{dubey2024llama} for text-only models and MMMU-pro~\cite{yue2024mmmu} for multi-modality models. MMLU-pro's maximum length is only 3076, which is shorter than the sliding window size of \gemma and Ministral. These two models will be degenerated into self-attention-only models with MMLU-pro dataset. Therefore, we use arXiv-QA, a long-context dataset that do question answering on a collection of arXiv articles~\cite{arxivdataset} for the two models.


\tightsubsection{End-to-end Evaluation}


\para{End-to-end throughput} \Cref{fig:eva-e2e-throughput} compares the end-to-end inference throughput of vLLM and \sys on both H100 and L4 GPUs. \sys achieves up to 4.92$\times$ speedup (1.80$\times$ on average) on H100 and 3.29$\times$ speedup (1.69$\times$ on average) on L4. The speedup comes from both less memory waste and better prefix caching. We will provide more breakdowns in \Cref{sec:ablation}. \sys also achieves comparable throughput with vLLM on the standard \llama, proving that it can also be used to serve self-attention-only models without introducing new overhead.

The smallest Jamba model is 52B, which cannot be filled in one L4 24GB GPU, so this model is skipped on the L4 platform. vLLM fails to serve the longest request in the dataset for the Ministral model on the L4 platform, while \sys can serve it due to the reduced memory usage. The throughput of Ministral is much smaller than other models because the requests have an average length of 92408, much longer than the thousands-of-token requests in other models.

\para{End-to-end latency} \Cref{fig:eva-e2e-latency} shows the latency under different request rates of the Llama Vision model. When the request rate is low ($<$1.2 requests/s), the latency of vLLM and \sys is similar, with only a 2.6\% difference on average, proving that \sys does not sacrifice model latency. When the request rate grows, \sys reduces the end-to-end latency by up to 2.24$\times$ and time to first token by up to 29.43$\times$ due to less memory waste and larger batch size. The time per output token (TPOT) of \sys is larger than vLLM because \sys batches more requests and has more computation in each step. \sys can achieve the same TPOT if scheduling the same number of requests in each step.


\tightsubsection{Break down}

\begin{figure}[t]
    \centering
    \includegraphics[width=.7\columnwidth]{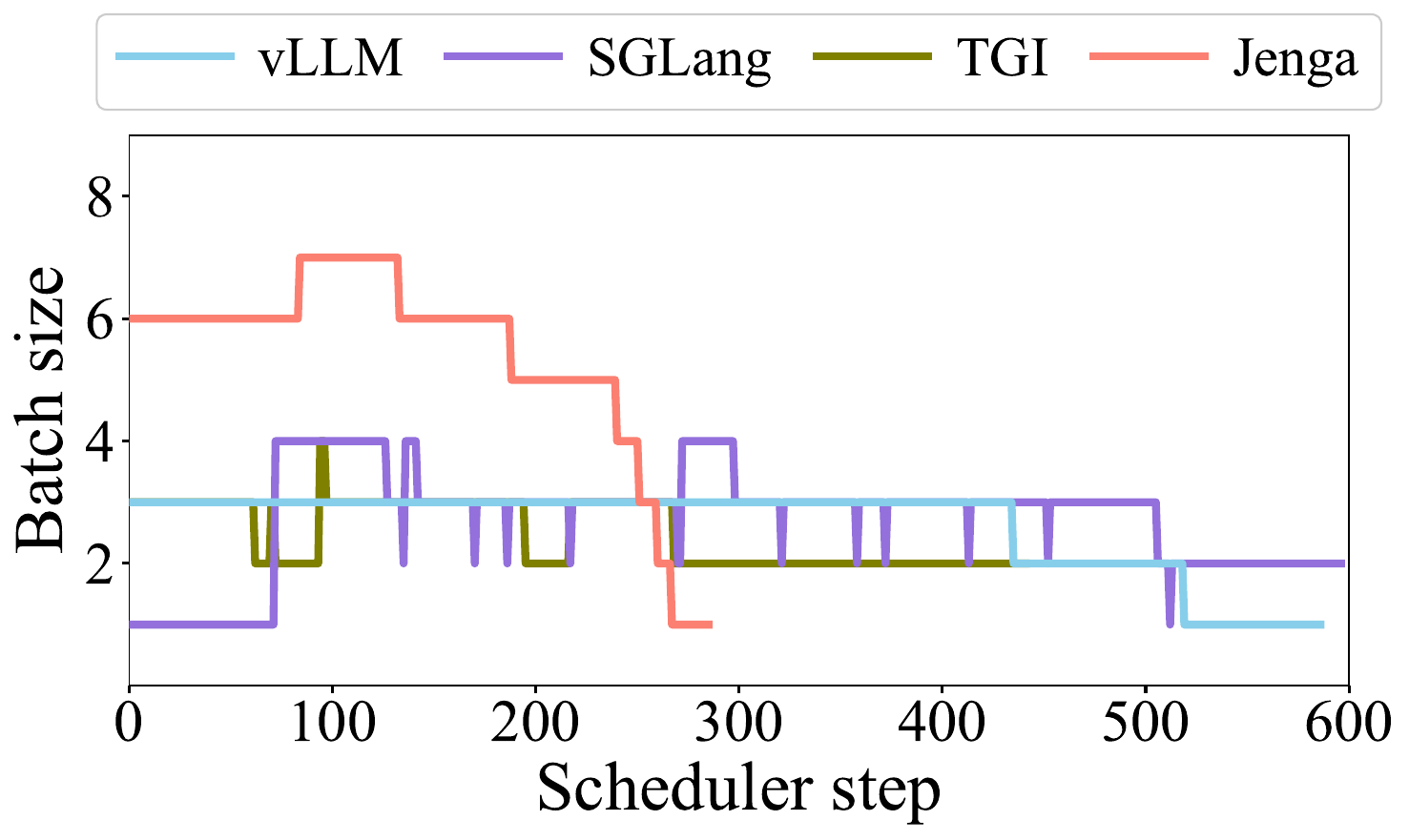}
    \vspace{-1em}
    \caption{Timeline of decode batch size for Ministral model of \sys and existing LLM inference engines.}
    \label{fig:eva-ministral-batchsize}
\end{figure}

\label{sec:ablation}
\para{Decode batch size} \sys improves the LLM serving engine's throughput and end-to-end latency by maximally enlarging the batch size. 
To measure the batch size of \sys compared to other inference engines, we pick three open-source inference engines (vLLM~\cite{kwon2023efficient}, SGLang~\cite{zheng2023efficiently} and TGI~\cite{tgi}) that are widely used in production. As for the workload, we use a simulated workload, where there are 20 requests coming to the inference engine all at once, with input length randomly drawn from 55-110 thousand tokens and with output length from 50-100 tokens. This simulates the typical long document question-answering workload, where the input length can be excessively long, but the output length is relatively short.
We then visualize the batch size of LLM decoding steps in \Cref{fig:eva-ministral-batchsize}.
The average batch size of \sys is 5.39, 1.95$\times$ larger than the average batch size of other inference engines (2.63, 2.74, and 2.50 for vLLM, SGLang, and TGI, respectively). As a result, \sys finishes LLM inference within 300 steps, while other inference engines need around 600 steps. Note that TGI finishes earlier as it does not support the \verb|--ignore-eos| flag~\cite{vllmblogpost}, and thus will generate fewer tokens compared to the other inference engines.

\begin{figure}[t]
\includegraphics[width=\columnwidth]{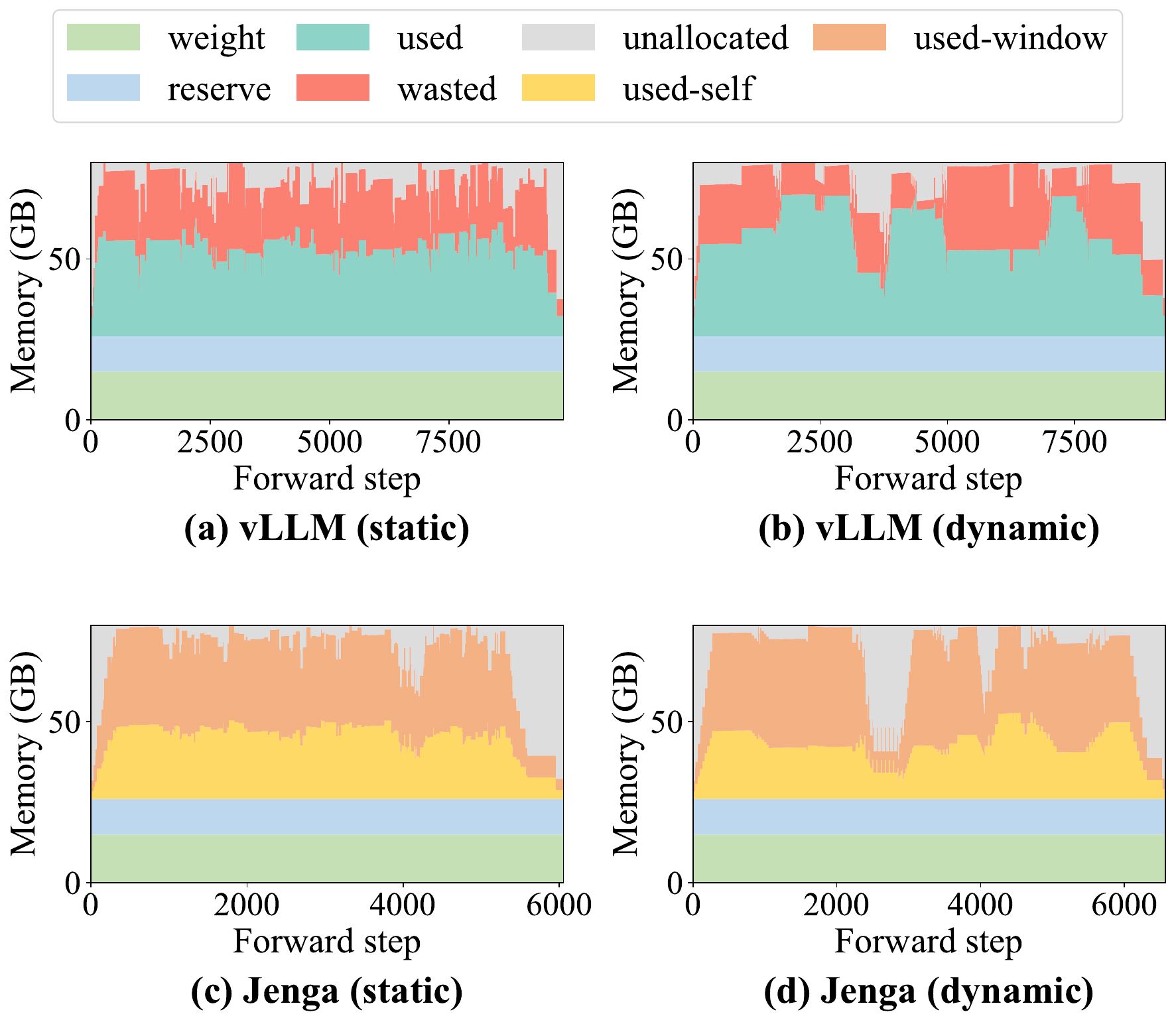}
\vspace{-2em}
\caption{Timeline of memory usage for Ministral model. vLLM shows significant wasted memory due to memory fragmentation (in red), while \sys minimizes waste.}
\label{fig:eva-memory-timeline}
\end{figure}

\para{Fragmentation analysis} \Cref{fig:eva-memory-timeline} shows the memory used for each part during the inference of Ministral model. We use a \textit{static} trace that the request length distribution does not change over time and a \textit{dynamic} trace that the average length forms a uniform distribution over time. We divide the use of GPU memory into five types, (1) the model \textit{weight}, (2) the memory \textit{reserved} for the inference engine for things like model activations and cuda graphs, (3) the memory used for storing KV caches required by new token generation (\textit{used} in \Cref{fig:eva-memory-timeline}\cgreen{ab}, \textit{used-self} and \textit{used-window} in \Cref{fig:eva-memory-timeline}\cgreen{cd}), (4) the \textit{wasted} memory that is allocated but not stores useful KV cache, and (5) the \textit{unallocated} KV cache memory. For the two traces, vLLM wastes 38.2\% KV cache memory on average as it fails to free the KV cache of sliding window layers for tokens outside the window, while \sys only has $0.04\%$ KV cache memory waste, which comes from the unused small pages inside the large pages and the last page that is only partially filled. Moreover, in the dynamic trace, \sys can dynamically allocate the KV cache memory to sliding window layers and self-attention layers based on the workload, with the rate of self-attention KV cache ranging from 27.8\% to 54.5\% among all allocated KV cache memory.

\begin{figure}[t]
\begin{subfigure}{.23\textwidth}
    \centering
    \includegraphics[width=.9\linewidth]{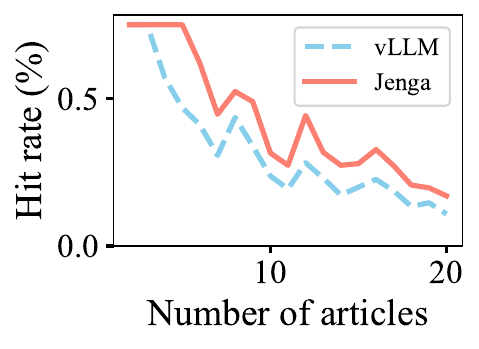}
    \vspace{-0.8em}
    \caption{Hit rate}
    \label{fig:eva-prefix-hit}
\end{subfigure}
\begin{subfigure}{.23\textwidth}
    \centering
    \includegraphics[width=\linewidth]{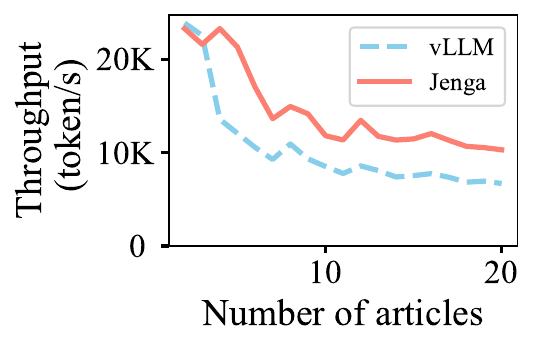}
    \vspace{-2em}
    \caption{Throughput}
    \label{fig:eva-prefix-throughput}
\end{subfigure}
\caption{Prefix caching with different number of articles.}
\label{fig:eva-prefix}
\end{figure}

\vspace{-0.1em}
\para{Prefix caching} \Cref{fig:eva-prefix} evaluates the prefix caching system of \sys by using a different number of articles in the arXiv dataset~\cite{arxivdataset} and asking multiple questions at the end of each article. When the number of articles is small (e.g., $<=3$, both the two systems can cache all the articles and provide similar throughput. The reason for the slight overhead of \sys is that \sys needs to allocate memory twice, one for self-attention layers and the other for sliding window layers, while vLLM only allocates once for all layers. When the number of articles is big, \sys has up to 1.60$\times$ higher cache hit rate as the customized sliding window eviction rule prioritizes the eviction of KV cache for tokens outside the sliding window, while vLLM treats all layers as self-attention layers. The higher cache hit rate saves more computation and thus provides up to 1.77$\times$ throughput improvement.

\begin{figure}[t]
\begin{subfigure}{.23\textwidth}
    \centering
    \includegraphics[width=\linewidth]{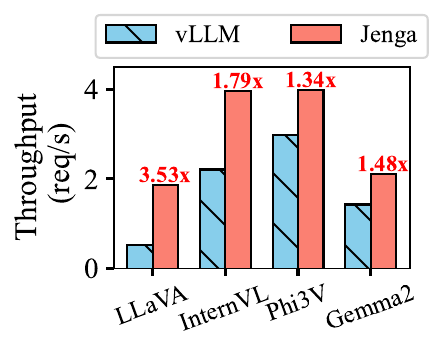}
    \vspace{-2em}
    \caption{Throughput}
    \label{fig:eva-mm-throughput}
\end{subfigure}
\begin{subfigure}{.23\textwidth}
    \centering
    \includegraphics[width=\linewidth]{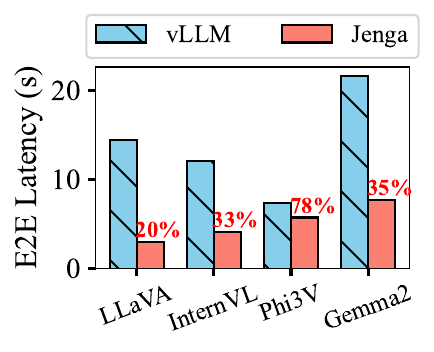}
    \vspace{-2em}
    \caption{E2E latency}
    \label{fig:eva-mm-itl}
\end{subfigure}
\caption{Vision language model with chunked prefill.}
\label{fig:eva-mm}
\end{figure}

\tightsubsection{Case Study}
\para{Vision embedding cache for VLMs} \Cref{fig:eva-mm} shows the performance improvement of vision language models due to \sys's support of vision embedding cache. Without such a cache, inference engines such as vLLM and SGLang need to re-run the vision encoder part in each chunked prefill step. With the vision embedding cache, \sys only needs to run the vision encoder once for each request, leading to 1.88$\times$ and 1.60$\times$ improvement in throughput and latency over vLLM, respectively. The models are evaluated on the MMMU-pro dataset with chunked prefill batch size 1024.

\begin{figure}[t]
    \centering
    \includegraphics[width=0.45\textwidth]{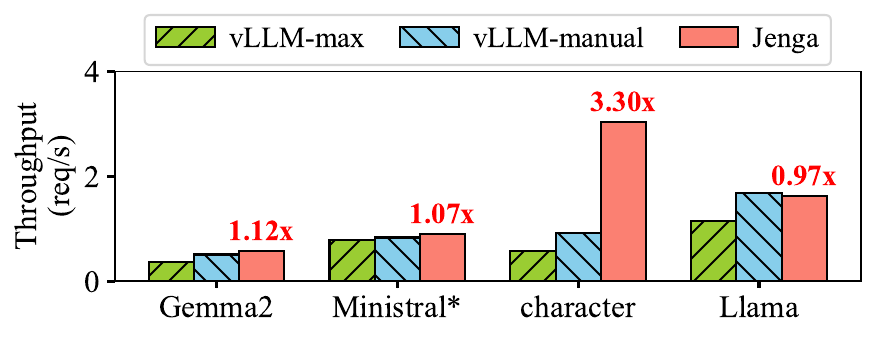}
    \vspace{-0.8em}
    \caption{Speculative decoding. Amplified the throughput of Ministral by 10 $\times$ for better visualization.}
    \label{fig:eva-spec-decode}
\end{figure}

\para{Speculative decoding} \Cref{fig:eva-spec-decode} compares the performance of speculative decoding in vLLM and \sys, which includes a small model and a large model running simultaneously. The large models use the model size in \Cref{tab:eva-setup}, and the small model sizes are 2B for Gemma2 and 1B for other models, where the 1B model of ministral is an example model created by us following the model configuration of \llama 3.2 1B. 

\textit{vLLM-max} refers to the scenario of using a uniform page size as in the PagedAttention~\cite{vllmblogpost}, where the page size needs to be set as the page size of the large model. \textit{vLLM-manual} uses a manually-designed memory allocation strategy for speculative decoding by SmartSpec~\cite{liu2024optimizing}. This strategy has no memory fragmentation when the model only contains self-attention layers but does not work perfectly on heterogeneous LLMs.  \sys can achieve the same performance as \textit{vLLM-manual} in standard \llama, showing the automatic memory management in \sys can reach the optimal case for self-attention-only models. Moreover, \sys can achieve an averaged 1.58$\times$ throughput improvement on heterogeneous LLMs without redesigning the memory allocation strategy.





 



\tightsection{Related Work}

\para{LLM serving systems} Many works have optimized LLM serving for different scenarios. Orca~\cite{yu2022orca} enables token-level request batching to enhance inference throughput. LoongServe~\cite{wu2024loongserve} and  DeepSpeed-FastGen~\cite{holmes2024deepspeed} improve the efficiency of long-sequence inference through advanced scheduling algorithms. DistServe~\cite{zhong2024distserve} and SARATHI~\cite{agrawal2024taming} mitigate output token latency variance by controlling the number of tokens processed per step. Parrot~\cite{lin2024parrot}, SGLang~\cite{zheng2023efficiently}, and XGrammar~\cite{dong2024xgrammar} enable users to specify the output structure. MuxServe~\cite{duan2024muxserve} allows multiple models to be served on a single GPU. These approaches have successfully optimized serving for homogeneous models with standard self-attention layers, and can be further extended to support heterogeneous models with the help of \sys.

\para{LLM Memory allocation} Various methods have been proposed to reduce memory fragmentation during LLM inference. PagedAttention~\cite{kwon2023efficient} eliminates fragmentation caused by variable-length sequences by dividing the KV cache into fixed-size pages, but fails to handle heterogeneous models with different embedding sizes. vAttention~\cite{prabhu2024vattention} utilizes GPU virtual memory to allocate contiguous KV cache memory for each request. However, it suffers from coarse-grained memory allocation and significant allocation and deallocation overhead of GPU driver. Moreover, virtual-memory-based mechanisms cannot track the prefix-subset dependency to perform effective prefix caching. Several studies propose memory allocation algorithms for specific model type, e.g., SLoRA~\cite{sheng2024slora} for LORA, Marconi~\cite{pan2024marconi} for Mamba, and SmartSpec~\cite{liu2024optimizing} for speculative decoding. In contrast, \sys provides a general solution that is compatible to a wider range of LLM memory types.

\para{LLM Memory optimization} There are also works to reduce the peak GPU memory usage of LLM inference. FlashAttention~\cite{dao2022flashattention,dao2023flashattention,shah2024flashattention} reduces the memory footprint of attention kernels by tiling. CachedAttention~\cite{gao2024cost} and Mooncake~\cite{qin2024kimi} enable KV cache offloading to larger memory pools, such as CPU memory or disk storage. \sys can provide fixed-size offloading granularity and suggest the offload order of pages when extending these systems to heterogeneous LLM senarios. Solutions such as FlexGen~\cite{sheng2023flexgen}, MoE-Lightning~\cite{cao2024moe}, Moe-Infinity~\cite{xue2024moe}, and PowerInfer~\cite{song2024powerinfer} further support weight and activation offloading. Additionally, various attention mechanisms have been developed to reduce KV cache size, including MQA~\cite{shazeer2019fast}, GQA~\cite{ainslie2023gqa}, and MLA~\cite{liu2024deepseek}, which decrease KV memory requirements per token, and several KV cache pruning techniques~\cite{xiao2023efficient,yang2024pyramidinfer,cai2024pyramidkv,adnan2024keyformer} to reduce the number of tokens inside the KV cache. These novel KV cache designs can be easily integrated into inference engines with the help of \sys.

\tightsection{Conclusion}

This paper introduces \sys, an efficient memory allocation framework for managing heterogeneous embeddings in modern LLM architectures.
By utilizing a two-level memory allocator, \sys reduces memory fragmentation and enables customizable caching policies for different types of embeddings.
In our experiments, \sys achieved up to 79.6\% higher GPU memory utilization and 1.26 -- 4.91$\times$ increased serving throughput across a variety of models and scenarios.

\bibliographystyle{plain}
\bibliography{refs}

\end{document}